\documentclass[twocolumn,showpacs,preprintnumbers,amsmath,amssymb]{revtex4}

\usepackage{graphicx}
\usepackage{bm}

\begin{document}

\title{Generalized van der Waals theory of liquid-liquid phase transitions}

\author{Yu. D. Fomin}
\affiliation{Institute for High Pressure Physics, Russian Academy
of Sciences, Troitsk 142190, Moscow Region, Russia}

\author{V. N. Ryzhov}
\affiliation{Institute for High Pressure Physics, Russian Academy
of Sciences, Troitsk 142190, Moscow Region, Russia}

\author{E. E. Tareyeva}
\affiliation{Institute for High Pressure Physics, Russian Academy
of Sciences, Troitsk 142190, Moscow Region, Russia}

\date{\today}

\begin{abstract}
In the framework of the thermodynamic perturbation theory for
fluids we study how the phase diagram of an isotropic repulsive
soft-core attractive potential, where a liquid-liquid phase
transition exists in addition to the standard gas-liquid phase
transition, changes by varying the parameters of the potential. We
show that existence of the liquid-liquid transition is determined
by the interplay of the parameters of the potential and the
structure of a reference liquid.
\end{abstract}

\maketitle

Despite the growing interest to the possible polymorphic phase
transitions in liquids and glasses \cite{book1,bl} the nature of
different phases which can be found in dense (and possibly
metastable) liquids is still puzzling. The coexistence of
different phases, while common for mixtures, is unexpected for a
simple fluid. In principle, the rules of thermodynamics do not
forbid the existence of more than two different fluid phases in
a simple fluid, however, from a common experience only two of
them are well known: a low density fluid (vapor) and a high
density fluid (liquid) phases. At the same time  in recent years
experimental evidences of such features of phase diagram as
liqiud-liquid transitions, polyamorphism, etc appeared for a
wide range of systems including water, $Si, I, Se, S, P$, etc
\cite{book1,bl,angell1,brazh,stanley1,mishima,thomp,brazh1,brazh2,togaya,katayama}.
The complexity of the phase diagrams in these substances may be
a result of complex interactions depending on the intermolecular
orientations.

At the same time exploring the possibility that simple fluids
interacting through isotropic potentials may exhibit the similar
behavior represents a serious challenge for theorists.

It was shown recently through molecular dynamics simulations
that a system of particles with the isotropic repulsive
soft-core attractive potential may have high-density and
low-density liquid phases \cite{stanley4,stan-etal}. This
potential may be considered as an effective potential resulting
from an average over the angular degrees of freedom for systems
where the position of the minimum approach between particles
depends on their relative orientations like in the case of the
hydrogen bonding between water molecules
\cite{book1,stanley1,water1,book2}. This potential may be also
used to model interactions in a variety of systems including
liquid metals, colloids, silica \cite{book1,book2}.

After the pioneering work by Hemmer and Stell \cite{stell}, where
the soft core potential with an attractive interaction at large
distances was first proposed for the qualitative explanation of
the solid-solid critical point in materials such as Ce or Cs, a
lot of attention was paid to the investigation of the properties
of the systems with the potentials that have a region of negative
curvature in their repulsive core. In spite of the simplicity of
the model, the physical mechanism that causes the liquid-liquid
phase transition in such systems is not completely understood. As
was emphasized in \cite{st_new1,st_new2} it arises from an
interplay of the different components of the pair interaction. In
Refs. \cite{st_new1,st_new2} authors tried to disentangle the role
of each component to investigate the dependence of the phase
diagram on the potential parameters. In Ref. \cite{st_new1} the
results of molecular dynamics calculations performed for several
sets of parameters were presented. The resulting behavior of the
critical points was interpreted through a modified van der Waals
equation where the effect of the repulsive shoulder at different
densities $\rho$ and temperatures $T$ can be taken into account by
an effective excluded volume depending on both $\rho$ and $T$. In
Ref. \cite{st_new2} the same analysis was undertaken by using an
integral equation approach in the hypernetted-chain approximation.
In Refs. \cite{st_new1,st_new2} it was shown that the high-density
liquid-liquid critical point can be found only when there is some
kind of balance between the attractive and repulsive parts of the
potential.

\begin{figure}
\includegraphics{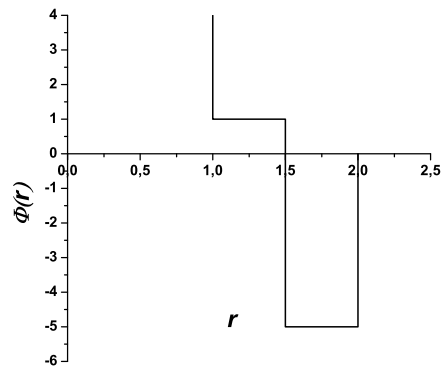}%
\caption{\label{fig:fig1} An isotropic potential with an
attractive well and a repulsive component consisting of a hard
core plus a finite shoulder. }
\end{figure}

It should be noted that it is widely believed (see, for example,
\cite{jagla1,jagla2}) that a fluid-fluid transition should be
related to the attractive part of the potential. However, in
Ref. \cite{RS2003} in the framework of the generalized van der
Waals theory it was shown that the purely repulsive step
potential is sufficient to explain the high density
liquid-liquid phase transition. In the present paper we apply
this theory to the study of the phase diagram of the system of
particles interacting through an isotropic potential with an
attractive well and a repulsive component consisting of a hard
core plus a finite shoulder. This potential can be written in
the form:

\begin{equation}
\Phi (r)=\left\{
\begin{array}{lll}
\infty , & r\leq d \\
\varepsilon_1 , & d <r\leq \sigma  \\
\varepsilon_2 , & \sigma <r\leq \sigma _{1} \\
0, & r>\sigma _{1}%
\end{array}%
\right. . \label{1}
\end{equation}

The potential is shown in Fig. (\ref{fig:fig1}). We apply to the
problem the first order thermodynamic perturbation theory for
fluids. The soft core of the potential (\ref{1})  is treated as
perturbation with respect to the hard sphere potential. In this
case the free energy of the system may be written in the form
\cite{barhen1,barhen2}:
\begin{gather}
\frac{F-F_{HS}}{Nk_{B}T} =\frac{1}{2}\rho \beta \int u_{1}(r)g_{HS}(r)d%
\mathbf{r},  \label{2}
\end{gather}%
where $\rho =V/N$ is the mean number density, $\beta =1/k_{B}T$,
$\ u_{1}(r)$ is the perturbation part of the potential
$u_{1}(r)=\Phi (r)-\Phi _{HS}(r)$, $\Phi _{HS}(r)$ is the hard
sphere singular potential, $g_{HS}(r)$ is the hard sphere radial
distribution function, which is taken in the Percus-Yevick
approximation \cite{henderson}.

To calculate $F_{HS}$, one can use, for example, the approximate equation %
\cite{barhen2}:
\begin{equation}
\frac{F_{HS}}{k_BTN}=3\ln\lambda-1+\ln\rho+\frac{4\eta-3\eta^2}{(1-\eta)^2}.
\label{3}
\end{equation}
Here $\lambda=h/(2\pi mk_BT)^{1/2}$ and $\eta=\pi\rho\sigma^3/6$.

\begin{figure}
\includegraphics{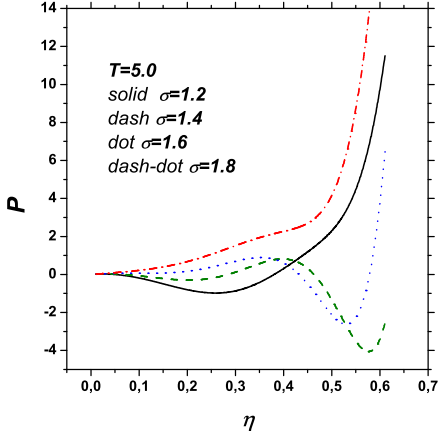}%
\caption{\label{fig:fig2} The pressure isotherms for the system
with $\varepsilon / \varepsilon_1=5$ for different values of
$\sigma/d$ for $T=5$ ($\sigma_1=2$).}
\end{figure}

Further in this paper we use the dimensionless quantities:
$\tilde{{\bf r}}={\bf r}/d$, $\tilde{P}=P\sigma
^{3}/\varepsilon ,$ $\tilde{V}=V/N\sigma ^{3}=1/\tilde{\rho},$ $\tilde{T}%
=k_{B}T/\varepsilon $, omitting the tilde marks.

\begin{figure}
\includegraphics[width=7cm]{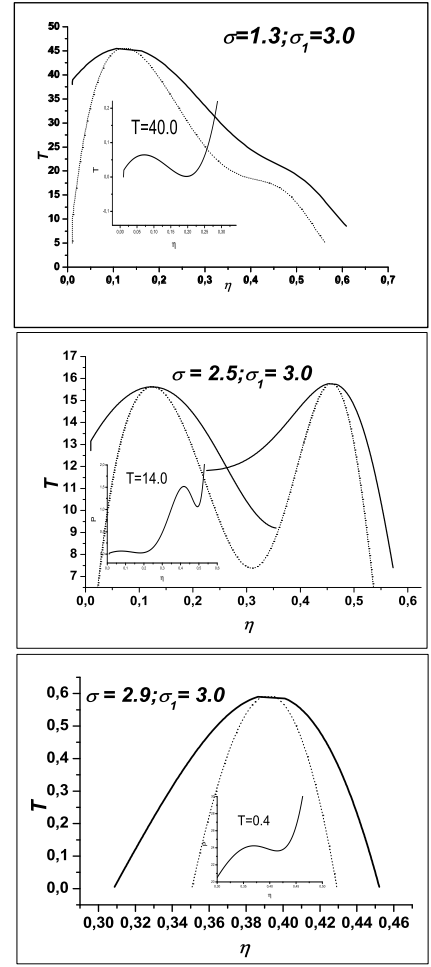}%
\caption{\label{fig:fig3} Phase diagram of the system of particles
interacting through the potential (1) for different values of
$\sigma / d $ and fixed value of $\sigma_1/d=3$. Solid lines
correspond to binodals and the dashed lines - to spinodals.
Inserts show the characteristic isotherms.}
\end{figure}

\begin{figure}
\includegraphics[width=8cm]{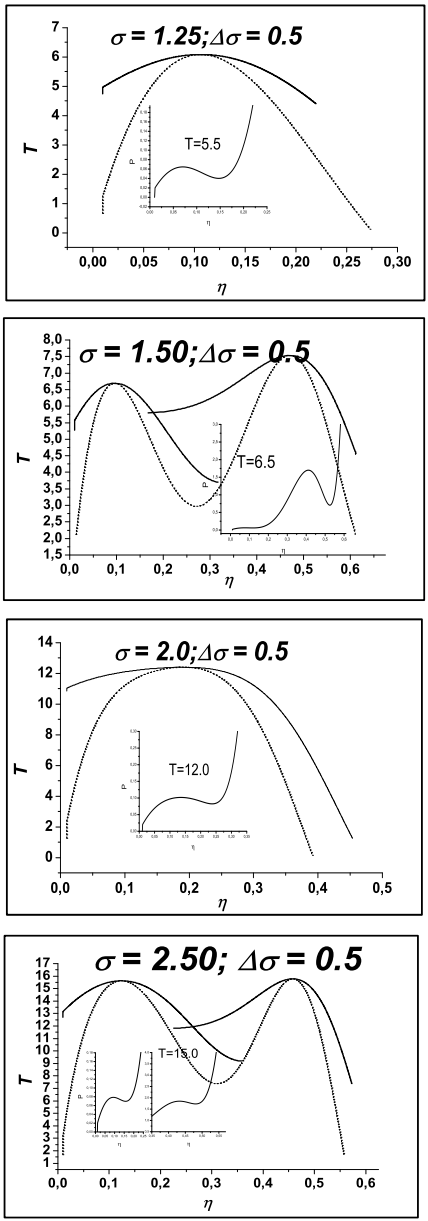}%
\caption{\label{fig:fig4} Phase diagram of the system of particles
interacting through the potential (1) for different values of
$\sigma / d $ and fixed value of
$\Delta\sigma=\sigma_1/d-\sigma/d=0.5$. Solid lines correspond to
binodals and the dashed lines - to spinodals. Inserts show the
characteristic isotherms.}
\end{figure}

Results of calculations are demonstrated in
Figs.~\ref{fig:fig2}-\ref{fig:fig4}. In Fig.~\ref{fig:fig2} a
family of pressure isotherms is shown for the system with
$\varepsilon / \varepsilon_1=5$ for different values of
$\sigma/d$ for $T=5$. The van der Waals loops in the isotherms
at low temperatures are clearly seen, this indicates the
existence of the first order liquid-gas and liquid-liquid phase
transitions.

Using the Maxwell construction we are able to calculate the
equilibrium lines of the liquid-liquid phase transitions at
different values of $\sigma / d $
(Figs.~\ref{fig:fig3}-\ref{fig:fig4}) (binodals) which are shown
as solid lines in the figures. The dashed lines correspond to
spinodals which as usual are calculated from the condition
$\partial P/\partial \rho = 0$. In Fig. \ref{fig:fig3} the
evolution of the phase diagram is shown as a function of $\sigma /
d $ for the fixed value of $\sigma_1/d=3$. We can see that the
phase diagram starts from one gas-liquid transition for $\sigma /
d = 1.3$. If value of $\sigma / d $ increases the second
(liquid-liquid) transition develops. When difference $\sigma_1 / d
- \sigma / d $ is small enough ($\approx 0.2$) the gas-liquid
transition disappears, and one has the only liquid-liquid
transition, as was discussed in Ref.~\cite{RS2003}. We cannot
extend the transition lines down to zero temperature and make
calculations for $\sigma_1/d <1.3$ because of limitation of the
perturbation approach.

\begin{figure}
\includegraphics{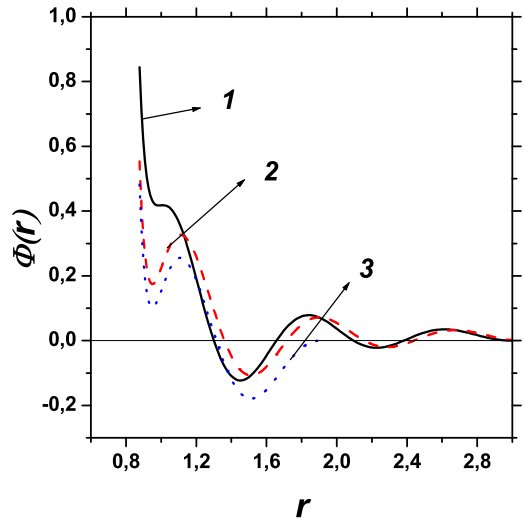}%
\caption{\label{fig:fig5} The potential (\ref{pot2}) for
$r_0=1.1$ (line 1) and $r_0=1.16$ (line 2). The line 3 is
obtained from the line 2 by the shift downward on the height of
the first maximum.}
\end{figure}

Fig.~\ref{fig:fig4} shows the evolution of the phase diagram of
the system of particles interacting through the potential (1) for
different values of $\sigma / d $ and fixed value of the
difference $\Delta\sigma=\sigma_1/d-\sigma/d=0.5$. It should be
noted that in contrast to Fig.~\ref{fig:fig3} the liquid-liquid
phase transition disappear for values of $\sigma$ in the vicinity
of $\sigma\approx 2$. This seems to contradict the assertion in
Refs. \cite{st_new1,st_new2} that the existence of a liquid-liquid
phase transition is simply determined by the some kind of a
balance between the repulsive and attractive parts of the
potential, but depends also on the positions of the maxima of a
reference radial distribution function (or the structure of the
reference liquid). This is seen from the right hand side term of
Eq. (\ref{1}).

To illustrate the application of the generalized van der Waals
theory to other systems let us consider the potential of the form:
\begin{equation}
\Phi (r)=\frac{a \exp(-\alpha r)\cos(2
k_f(r-r_0))}{r^3}+b\left(\frac{\sigma}{r}\right)^{18},
\label{pot2}
\end{equation}
where $\alpha=0.1$, $a=0.5$, $k_f=4.1$, $\sigma=0.331$,
$b=0.42\times10^8$. This potential may be used, for example, for a
qualitative modelling of effective potentials of some metals
\cite{met_pot}. In Fig.~\ref{fig:fig5} this potential is shown for
$r_0=1.1$ (line 1) and $r_0=1.16$ (line 2). The line 3 is obtained
from the line 2 by the shift downward on the height of the first
maximum.

\begin{figure}
\includegraphics[width=8cm]{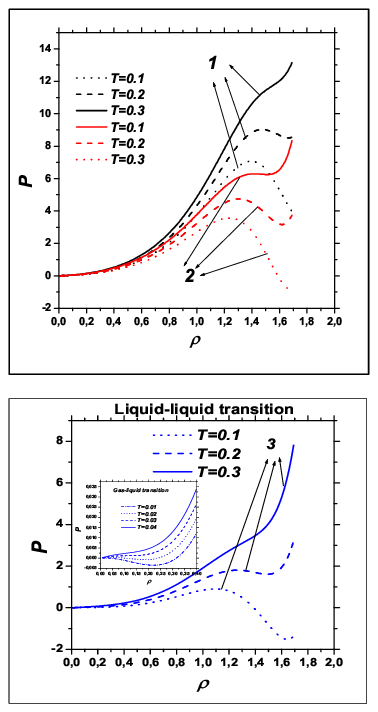}%
\caption{\label{fig:fig6} The families of isotherms
corresponding to the potentials in Fig.~\ref{fig:fig5}. There
are only liquid-liquid transitions for the potentials
corresponding to the lines (1)and (2) in Fig.~\ref{fig:fig5}
(upper figure). For the potential corresponding to the line (3)
in Fig.~\ref{fig:fig5} there is also a low-temperature
gas-liquid transition shown in the insert in the lower figure.}
\end{figure}

In Fig. \ref{fig:fig6} the corresponding families of isotherms
are shown for different temperatures. To calculate the equation
of state corresponding to the potential (\ref{pot2}) we apply
the thermodynamic perturbation theory. The second part of the
potential (\ref{pot2}) was considered as a reference system.
\begin{equation}
\Phi_0(r)=b\left(\frac{\sigma}{r}\right)^{18}. \label{pot0}
\end{equation}
Potential $\Phi_0(r)$ is approximated by the hard sphere
potential with an effective diameter which depends on the
density and the temperature \cite{VW}.

From Fig.~\ref{fig:fig6} one can see that there are only
high-density liquid-liquid transitions for the potentials
corresponding to the lines (1)and (2) in Fig.~\ref{fig:fig5}
(upper figure). These transitions exist due to negative
curvature of the potential (\ref{pot2}). There are no gas-liquid
transitions in these cases because the strength of an attraction
is insufficient. For the potential corresponding to the line (3)
in Fig.~\ref{fig:fig5} there is also a low-temperature
gas-liquid transition shown in the insert in the lower figure.

We would like to emphasize that we do not claim that the first
order perturbation scheme, which was used in the present
article, gives the high precision quantitative results, however,
it seems reliable enough to give correct qualitative description
of the liquid-liquid transition in the system with the
potentials (\ref{1}) and (\ref{pot2}). It should be noted that
the second-order perturbation theory gives qualitatively the
same results.

Finally, let us make some remarks on another possible mechanism
of a liquid-liquid phase transition. The crystalline solid state
is characterized by a long-range positional order of the atomic
density and associated bond orientational order belonging to one
of the well-known allowed crystalline symmetries. In a
diffraction experiment, $\delta$-function Bragg reflections are
observed in the structure factor $S({\bf q})$ that shows the
associated point symmetry of the lattice. In the case of the
long range bond-orientational order, one can observe the
modulation of the structure factor which corresponds to the
symmetry of the nearest neighbor environment of a particle (see,
for example, \cite{ryzh91}).

However, one can imagine the situation when the isotropy of three
or four-particle correlation function is broken, but the symmetry
of one and two-particle correlation functions is unchanged. In
this case the structure factor of the system is the same as for
the isotropic liquid because it depends only on the two-particle
correlation function. This is why we call this type of the
symmetry a "hidden symmetry" - it can not be detected in
diffraction experiments.

So, using the diffraction experiments it is impossible to
distinguish a liquid-liquid transition without changing the
symmetry of the correlation functions and the transition
accompanied by the breaking of a hidden symmetry. It seems that
the only way to answer the question what kind of transition one
observes in an experiment is to analyze the whole phase diagram.
In principle the hidden symmetry transition may be of the first
order, but in contrast to the liquid-liquid transition discussed
in this article, it separates the phases with different symmetries
and can not end at the critical point. There are two possibilities
for the hidden symmetry phase transition:(1) it may end at the
tricritical point, and measurements above the tricritical point
should reveal the thermodynamic anomalies but no density change;
(2) it may intersect the gas-liquid transition line. In this case
one can expect a sharp bend on the gas-liquid transition line at
the point of intersection.

To our knowledge there are no calculations exploring the idea of
breaking of the symmetry of higher order distribution functions
in application to the problem of liquid-liquid transitions.

\begin{acknowledgments}
We thank S. M. Stishov, V. V. Brazhkin and A. G. Lyapin for
stimulating discussions. The work was supported in part by the
Russian Foundation for Basic Research (Grants No 05-02-17280 and
No 05-02-17621) and NWO-RFBR Grant No 047.016.001.
\end{acknowledgments}


\end{document}